\documentclass{article}
\usepackage{latexsym}
\usepackage{amscd}
\usepackage{amsmath}
\newcounter{saveeqn}%
\newcommand{\alphaeqn}{\setcounter{saveeqn}{\value{equation}}%
\stepcounter{saveeqn}\setcounter{equation}{0}%
\renewcommand{\theequation}
        {\mbox{\arabic{saveeqn}-\alph{equation}}}}%
\newcommand{\reseteqn}{\setcounter{equation}{\value{saveeqn}}%
\renewcommand{\theequation}{\arabic{equation}}}%

\begin{document}
\title{ Quasitemporal structure and symmetries in histories-based generalized quantum mechanics in curved spacetime}
\author{Tulsi Dass\thanks{Email: tulsi@iitk.ac.in}  and Yogesh Joglekar \thanks{Present address: Department of Physics, Indiana University, Bloomington, IN 47405} \footnote{email: yojoglek@indiana.edu } \\ Department of Physics \\ Indian Institute of Technology, Kanpur, India 208016}
\date{}
\maketitle


\begin{abstract}
The formalism for histories-based generalized quantum mechanics developed in two earlier papers is applied to the treatment of histories (of particles or fields or more general objects) in curved spacetimes (which need not admit foliation in spacelike hypersurfaces). The construction of the space of temporal supports (a partial semigroup generalizing the space of finite time sequences employed in traditional temporal description of histories) employs spacelike subsets of spacetime having dimensionality less than or equal to three. Definition of symmetry is sharpened by the requirement of continuity of mappings (employing topological partial semigroups). It is shown that with this proviso, a symmetry in our formalism implies a conformal isometry of the spacetime metric.
\end{abstract}


\section{INTRODUCTION AND BACKGROUND}
\label{secone}

Description of dynamics of closed systems (in particular, the universe) in terms of consistent histories [1-6] has been increasingly employed in the past one and a half decade to address fundamental questions in physics. In these theories time plays only a book keeping role (in the form of sequences $t_1,t_2,\cdots,t_n$, for example); its traditional use as a real variable in analytical work is absent. In curved spacetimes permitting foliation in spacelike hypersurfaces, the parameter labelling the leaves of foliation plays the role of time; discrete values of the same parameter can, of course, be employed for book-keeping of histories. (See, for example, Blencowe~\cite{mb}.)

Isham~\cite{cji} noticed that the general mathematical structure appropriate for the temporal description of histories is that of a partial semigroup (a set with an associative composition rule defined for some but not necessarily all pairs of elements). Given a history $\alpha=(\alpha_{s_1},\cdots,\alpha_{s_m})$ consisting of `events' $\alpha_{s_i}$ (projection operators in the appropriate quantum mechanical Hilbert space) at times $s_1,s_2,\ldots,s_m$ ($s_1<s_2<\cdots<s_m$), let us call the sequence $\xi=(s_1,s_2,\ldots,s_m)$ the temporal support of $\alpha$. Given another history $\beta=(\beta_{t_1},\cdots,\beta_{t_n})$ with temporal support $\eta=(t_1,\ldots,t_n)$ such that $s_m<t_1$ (we say that $\alpha$ temporally precedes $\beta$ and denote this as $\alpha\lhd\beta$), we can compose the two histories to obtain the history 
\begin{equation}
\label{one}
\gamma = \alpha\circ\beta=(\alpha_{s_1},\cdots,\alpha_{s_m},\beta_{t_1},\cdots,\beta_{t_n})
\end{equation}
whose temporal support is 
\begin{equation}
\label{two}
\zeta=\xi\circ\eta=(s_1,\ldots,s_m,t_1,\ldots,t_n).
\end{equation}
The composition rules (\ref{two}) and (\ref{one}) define the structure of partial semigroups on, respectively, the space $\mathcal{K}_1$ of temporal supports (the space of finite ordered sequences of real numbers) and the space $\mathcal{K}_2$ of homogeneous histories (\emph{i. e.} histories which can be represented as temporal sequences of `events'; one can think of more general histories like `$\alpha$ or $\beta$' which generally cannot be so represented). Elements of $\mathcal{K}_1$ and $\mathcal{K}_2$ admit irreducible decompositions of the form 
\alphaeqn
\begin{eqnarray}
\label{three}
\xi & = & {s_1}\circ{s_2}\circ\cdots\circ{s_m}, \\
\alpha & = & \alpha_{s_1}\circ\alpha_{s_2}\circ\cdots\circ\alpha_{s_m}. 
\end{eqnarray}
\reseteqn
The irreducible entities ${s_i}$ and $\alpha_{s_j}$ which do not admit further decomposition are called nuclear elements of $\mathcal{K}_1$ and $\mathcal{K}_2$ respectively. There is a partial semigroup homomorphism $\sigma$ of $\mathcal{K}_2$ onto $\mathcal{K}_1$ given by 
\begin{eqnarray}
\label{four}
\sigma(\alpha)=\xi, & \sigma(\beta)=\eta, &\sigma(\alpha\circ\beta)=\sigma(\alpha)\circ\sigma(\beta).
\end{eqnarray}
The triple $(\mathcal{K}_2,\mathcal{K}_1,\sigma)$ is a prototype of a `quasitemporal structure' (a pair of partial semigroups with a homomorphism of the first onto the other). History theories admitting such a quasitemporal structure $( \mathcal{U}, \mathcal{T}, \sigma)$ are called quasitemporal theories~\cite{cji}.

Basic concepts relating to the treatment of histories of quantum fields in general spacetimes were developed by Hartle~\cite{jh2} and Sorkin~\cite{rds} who proposed to employ averages of fields over spacetime regions as basic observables. A temporal order can be defined on a family of spacetime regions which is causally consistent (which means that the family has no pair of regions in which one region intersects both the future and the past of the other). Appropriate families of such regions serve as `time points' in the description of field histories.

Isham~\cite{cji} translated these ideas in the language of partial semigroups leading to a quasitemporal structure $(\mathcal{U},\mathcal{T},\sigma)$ for such theories. (Our notations for the space of temporal supports and some related objects are different from those of Isham.) Given a spacetime $(M,g)$, he defined a temporal support as a collection of (four dimensional) `basic regions' with appropriate temporal relation defined in terms of light cones. In this context, a temporal relation $\prec$ plays an important role. Given subsets $A$ and $B$ of $M$, we say that $A\prec B$ ($A$ precedes $B$) if 
\begin{eqnarray}
\label{five}
A\cap B=\phi, & J^+(A)\cap B\neq\phi, & J^+(B)\cap A=\phi.
\end{eqnarray}
Here $J^+(A)$ is the causal future of $A$ (\emph{i. e.} the set of points in $M$ that can be reached from $A$ by future directed non-spacelike curves).

It should be noted that the temporal supports employed by Blencowe~\cite{mb} [finite ordered sequences of (subsets of) spacelike hypersurfaces] cannot be considered as a subclass of the temporal supports of Isham~\cite{cji} because the latter employ four-dimensional basic regions. While, from a theoretical point of view, the use of four-dimensional regions is more satisfying (because it makes provision for finite spatio-temporal localization of `events'), it is quite often convenient and useful to work with idealized `time points' (which means, in the present context, three or lower dimensional basic regions). The construction of the space of temporal supports given by us in the next section employs such idealized `time points' and can be used to generalize Blencowe's treatment of field histories to general spacetimes.
 
In the quasitemporal history theories proposed by Isham based on triples $(\mathcal{U},\mathcal{T},\sigma)$, histories were taken to be (multitime or more general) propositions. The space $\mathcal{U}$ of `history filters' (homogeneous histories) was proposed to be embedded in a larger space $\Omega$ (denoted as $\mathcal{UP}$ in ref.~\cite{cji} and ~\cite{cjinl}) of history propositions (inhomogeneous histories) which incorporated the concepts of `$\alpha$ and $\beta$', `$\alpha$ or $\beta$', `not $\alpha$' etc. Another object of fundamental importance in the theory was the space $\mathcal{D}$ of decoherence functionals which are complex valued functions defined on pairs of elements of $\Omega$ and satisfy four standard conditions [3-6] of hermiticity, positivity, additivity and normalization; they serve to define decoherence condition for histories and probabilities of histories in a decoherent set. Isham and Linden~\cite{cjinl} proposed a more general framework for history theories in which the basic entities were the pair $(\Omega,\mathcal{D})$ (with no reference to any underlying quasitemporal structure) in which $\Omega$ was assumed to be an orthoalgbra [a set of propositions incorporating the operations of partial order $\leq$ (coarse graining), disjointness $\perp$ (mutual exclusion), a disjoint join operation $\oplus$ (`or' operation for mutually exclusive propositions) and a few other features]. Quasitemporal theories were supposed to be a subclass of Isham-Linden type theories. 

The Isham-Linden formalism is not adequately equipped to bring out the full dynamics content of histories of a system in an autonomous framework. To remedy this deficiency, an axiomatic framework for histories-based theories was proposed by us~\cite{tdyj1} in which the basic ingradients were an Isham type triple $(\mathcal{U},\mathcal{T},\sigma)$ with the proviso that the space $\mathcal{U}_\tau = \sigma^{-1}(\tau)$ for nuclear $\tau$ (spaces of `single time' history propositions) have the structure of a logic~\cite{vsv}; the space $\Omega$ of history propositions was explicitly constructed and shown to be an orthoalgebra. With appropriate choice of logics, history versions of classical or quantum mechanics of systems can be realized as special cases of this formalism. The concept of logic permits the introduction of single `time' states and observables and of temporal evolution. Decoherence functionals were taken to be objects defined in terms of a given initial state and a law of  temporal evolution. Explicit construction of decoherence functionals was given for Hilbert space-based theories (which have history version of traditional quantum mechanics as a special case) and for a somewhat generalized form of classical mechanics. The formalism of~\cite{tdyj1} will be referred to as the `augmented temporal logic formalism'.

In a recent paper~\cite{tdyj2} we have given a systematic treatment of symmetries and conservation laws for histories-based theories giving special consideration to the formalism of~\cite{tdyj1}. In this work symmetries were defined in terms of mappings on the spaces $\mathcal{U}$ and $\mathcal{T}$ which preserve the quasitemporal structure, logic structure of $\mathcal{U}_\tau$'s and an invariance condition involving decoherence functionals (which ensures preservation of the decoherence condition for histories and probabilities of histories in decoherent sets). Concepts of orthochronous (`temporal order' preserving) and non-orthochronous (`temporal order' reversing) symmetries were introduced. A simple criterion for physical equivalence of histories was introduced in terms of their being related through orthochronous symmetries; this criterion incorporated the various notions of physical equivalence of histories introduced by Gell-Mann and Hartle~\cite{mgjbh} as special cases. 

In the present work we apply the formalism of~\cite{tdyj1} and~\cite{tdyj2} to the treatment of histories-based dynamics (of general closed systems -- particles, fields, strings or more general objects) in curved spacetimes (which need not admit foliation in spacelike hypersurfaces) paying special attention to quasitemporal structure and symmetries. The quasitemporal structure is chosen so as to conform to the convention of~\cite{tdyj1}. This necessitates, firstly, the replacement of four dimensional basic regions of Isham by three (and lower) dimensional basic regions and, secondly, a modification of the temporal order relation (\ref{five}) of Isham [see eq. (\ref{six}) below]. In the treatment of symmetries, we impose the condition of continuity on the mappings mentioned above; this is done by taking the spaces $\mathcal{U}$ and $\mathcal{T}$ to be topological partial semigroups. We concentrate on the mapping $\Phi_1:\mathcal{T}\rightarrow\mathcal{T}$ (which is a part of definition of a symmetry), construct a topology for $\mathcal{T}$ explicitly (in an intuitively expected manner) and show that, with the continuity requirement imposed, the mapping $\Phi_1$ induces a transformation of the spacetime $M$ which is a conformal isometry of the metric $g$.

The rest of the paper is organized as follows. In section \ref{sectwo}, we present a development of histories based dynamics in curved spacetime along the lines of~\cite{tdyj1}. Main efforts are directed towards evolving a detailed quasitemporal structure along the lines mentioned above. Symmetries in this framework are treated in section \ref{secthree} along the lines of~\cite{tdyj2} and the result about conformal isometries mentioned above is obtained. The last section contains some concluding remarks. An appendix is devoted to the construction of the topology for the space $\mathcal{T}$ mentioned above and related matters.


\section{AUGMENTED TEMPORAL LOGIC FORMALISM FOR HISTORIES-BASED DYNAMICS 
IN CURVED SPACE-TIME}
\label{sectwo} 

The basic ingradients in the formalism of~\cite{tdyj1} are the triple $(\mathcal{U},\mathcal{T},\sigma)$. We first consider the construction of the space $\mathcal{T}$ of temporal supports. 

The spacetime $(M,g)$ is assumed to be a 4-dimensional manifold $M$ equipped with a metric $g$ of Lorentzian signature $(- + + +)$. Given two subsets $A$ and $B$ of $M$, we shall say that $A$ temporally precedes $B$ $(A\lhd B)$ if 
\begin{eqnarray}
\label{six}
J^+(A)\cap B\neq\phi, & & \left[ J^+(B)-B\right]\cap A =\phi.
\end{eqnarray}
We have modified Isham's definition (\ref{five}) to allow the possibility $A = B$ or, more generally, $A\cap B\neq\phi$. This is in keeping with our convention (which facilitated analytical work in~\cite{tdyj1}) of allowing, for example, $s_m = t_1$ in eq.(\ref{two}) (in particular ${t_1}\circ {t_1}={t_1}$ in $\mathcal{K}_1$). Like the relation $\prec$ given by eq.(\ref{five}), the relation $\lhd$ is also not a partial order; in particular, $A\lhd B$ and $B\lhd C$ does not generally imply $A\lhd C$.

By a basic region we shall mean a connected subset of $M$ such that every pair of points in it has spacelike separation. Thus, a basic region can be a single point, a spacelike curve, a two-dimensional spacelike surface or a three dimensional spacelike hypersurface. This is in contrast to~\cite{cji} where the basic regions were taken to be four dimensional.

A nuclear temporal support is a collection
\begin{equation}
\label{seven}
\tau=\{B_1,B_2,\ldots\}
\end{equation}
of basic regions such that all the pairs $B_i$, $B_j$ have mutually spacelike separation. Given two nuclear temporal supports $\tau=\{B_1,B_2,\ldots\}$ and $\tau'= \{B'_1,B'_2,\ldots\}$, we say that $\tau\lhd\tau'$ if 
\begin{equation}
\label{eight}
\cup_i(B_i) \lhd \cup_j(B'_j).
\end{equation}
A temporal support 
\begin{equation}
\label{nine}
\xi=\{\ldots,\tau_1,\tau_2,\ldots\}
\end{equation}
is a countable ordered collection of nuclear temporal supports such that consecutive entries in $\tau$ are temporally ordered (\emph{i. e.} $\tau_j\lhd\tau_{j+1}$ for all $j$) and, moreover, the collection is at most semi-infinite which means that it has either an `earliest' member or a `latest' member or both. The family of temporal supports will be denoted as $\mathcal{T}$ and the subfamily of nuclear temporal supports as $\mathcal{N}(\mathcal{T})$. [Identifying $\tau\in\mathcal{N}(\mathcal{T})$ with $\{\tau\}\in\mathcal{T}$, $\mathcal{N}(\mathcal{T})$ is clearly a subfamily of $\mathcal{T}$.]

A partial semigroup (psg) structure can be defined on $\mathcal{T}$ as follows: Given two elements $\xi$ and $\eta$ of $\mathcal{T}$, we say that $\xi\lhd\eta$ if $\xi$ has a latest member $\tau_0$ and $\eta$ has an earliest member $\tau'_1$ (\emph{i. e.} $\xi=\{\ldots,\tau_{-1},\tau_0\}$ and $\eta=\{\tau'_{1},\tau'_{2},\ldots\}$) such that $\tau_0\lhd\tau'_1$. If the joint collection of nuclear temporal supports in $\xi$ and $\eta$ is at most semi-infinite, we define the composition $\xi\circ\eta$ as 
\begin{equation}
\label{ten}
\xi\circ\eta=\{\ldots,\tau_{-1},\tau_0,\tau'_1,\tau'_2,\ldots\}.
\end{equation}
We assume that
\begin{equation}
\label{eleven}
\{\tau\}\circ\{\tau\}=\{\tau\} \mbox{ for all } \tau\in\mathcal{N}(\mathcal{T}).
\end{equation}
This has the implication that the composition (\ref{ten}) is also defined if $\tau_0=\tau'_1$.

A general element $\xi=\{\ldots,\tau_1,\tau_2,\ldots\} \in\mathcal{T}$ admits an irreducible decomposition of the form
\alphaeqn 
\begin{equation}
\label{12a}  
\xi=\cdots\circ\{\tau_1\}\circ\{\tau_2\}\circ\cdots
\end{equation}
which we simply write as
\begin{equation}
\label{12b}
\xi=\cdots\circ\tau_1\circ\tau_2\circ\cdots.
\end{equation}
\reseteqn
This irreducible decomposition is unique modulo the trivial redundancy implied by the convention of (\ref{eleven}).

In the appendix we give a straightforward construction of a topology on $\mathcal{T}$ to make it a topological partial semigroup (\emph{i. e.} a psg $\mathcal{T}$ which is also a topological space with a topology such that the composition $\circ$, considered as mapping from a subset of $\mathcal{T}\times\mathcal{T}$ into $\mathcal{T}$ is continuous). This is needed for a continuity argument in the next section.

We next consider the construction of the space $\mathcal{U}$ of history filters. The first step in this construction is to associate, with each basic region $B$, a logic $\mathcal{U}_B$ such that, for two basic regions $B$ and $B'$ having spacelike separation, the logics $\mathcal{U}_B$ and $\mathcal{U}_{B'}$ are isomorphic $(\mathcal{U}_B\approx\mathcal{U}_{B'})$. Identifying isomorphic logics, we can now associate a logic $\mathcal{U}_\tau$ with an element $\tau\in\mathcal{N}(\mathcal{T})$. (It is just the logic associated with any of its basic regions.) Logics associated with two different nuclear temporal supports generally need not be isomorphic. [This generally was kept in~\cite{tdyj1} to give the formalism additional flexibility so as to make it applicable to systems (for example, the Universe) whose empirical characteristics may change with time.]

A history filter may now be defined as an assignment, to some element $\xi=\{\ldots,\tau_1,\tau_2,\ldots\}$ of $\mathcal{T}$, a collection $\alpha=\{\cdots,\alpha_1,\alpha_2,\cdots\}$ such that
\begin{itemize}
\item[(i)] the entries in $\alpha$ are in one-one correspondence with those in $\xi$ preserving order;
\item[(ii)] $\alpha_j \in \mathcal{U}_{\tau_j}$ for every $j$.
\end{itemize}
We define the map $\sigma : \mathcal{U}\rightarrow\mathcal{T}$ such that $\sigma(\alpha)=\xi$. A temporal order relation $\lhd$ and a psg structure can now be defined on $\mathcal{U}$ in a fairy obvious manner so as to make $\sigma$ a psg homomorphism. Indeed, given $\alpha=\{\cdots,\alpha_{-1},\alpha_0\}$ and $\beta=\{\beta_1,\beta_2,\cdots\}$ with $\sigma(\alpha)=\xi=\{\ldots,\tau_{-1},\tau_0\}$ and $\sigma(\beta)=\eta=\{\tau'_1,\tau'_2,\ldots\}$ we say that $\alpha\lhd\beta$ if $\xi\lhd\eta$ (which means $\tau_0\lhd\tau'_1$). Moreover, if $\xi\circ\eta$ is defined, we define
\begin{equation}
\label{13}
\alpha\circ\beta=\{\cdots,\alpha_{-1},\alpha_0,\beta_1,\beta_2,\cdots\}.
\end{equation}
Clearly
\begin{eqnarray}
\label{14}
\sigma(\alpha\circ\beta)= & \{\ldots,\tau_{-1},\tau_0,\tau'_1,\tau'_2,\ldots\} 
= \xi\circ\eta &=\sigma(\alpha)\circ\sigma(\beta).
\end{eqnarray}

In keeping with the convention (\ref{eleven}) for $\mathcal{T}$, we adopt a similar convention for the space of $\mathcal{U}$: Given $\alpha={\alpha_1}$ such that $\sigma(\alpha)={\tau_1}$ where $\tau_1\in\mathcal{N}(\mathcal{T})$, we stipulate that 
\begin{equation}
\label{15}
\{\alpha_1\}\circ \{\alpha_1\}=\{\alpha_1\}.
\end{equation}
The mapping $\sigma : \mathcal{U}\rightarrow\mathcal{T}$ is easily seen to be a psg homomorphism; it is, of course, onto (\emph{i. e.} every element of $\mathcal{T}$ is the image under $\sigma$ of some element of $\mathcal{U}$).

In the terminology of~\cite{tdyj1}, $\mathcal{U}$ and $\mathcal{T}$ are directed, special psg's. The triple $(\mathcal{U},\mathcal{T},\sigma)$ satisfies the axiom $A_1$ of \cite{tdyj1}. One can now invoke the other axioms and the developments in~\cite{tdyj1} can proceed in a straightforward manner [leading to, among other things, the space $\Omega$ of history propositions (`inhomogeneous histories') which is manifestly an orthoalgebra]. 
It should be mentioned that the axiom $A_2$ of~\cite{tdyj1}, which excludes histories with `closed time loops' would exclude spacetimes with closed timelike curves.

The logics employed above can be quite general - they can be Boolean logics associated with classical mechanics of particles and/or fields, standard quantum logics (typically the space $\mathcal{P}(\mathcal{H})$ of projection operators in a separable Hilbert space $H$) or more general logics. The formalism presented above can, therefore, be applied to the history versions of the classical or quantum mechanics of particles, fields, strings or more general objects.

Explicit construction of the decoherence functionals for the two special subclasses of theories mentioned in the introduction can be easily adopted in the corresponding subclasses of theories considered in the present section (\emph{i. e.} when $\mathcal{U}_\tau$'s are either families of projection operators in separable Hilbert spaces or those of Borel measurable subsets of phase spaces of classical systems); we shall, however, skip the details.


\section{SYMMETRIES}
\label{secthree}

Following the general idea~\cite{hw} of defining symmetries as automorphisms (structure preserving invertible mappings) of the appropriate mathematical framework, we defined in~\cite{tdyj2} a symmetry of an `augmented history system' $({\sf S})=(\mathcal{U},\mathcal{T},\sigma,\Omega,\mathcal{D})$ as a triple $\Phi=(\Phi_1,\Phi_2,\Phi_3)$ of invertible mappings $\Phi_1:\mathcal{T}\rightarrow\mathcal{T}$, $\Phi_2:\mathcal{U}\rightarrow\mathcal{U}$ and $\Phi_3:\mathcal{D}\rightarrow\mathcal{D}$ such that the following conditions are satisfied. 
\begin{enumerate}
\item[(i)] $\Phi_1$ is an automorphism or anti-automorphism of $\mathcal{T}$, \emph{i.e.} it satisfies either (a) or (b) below.
\begin{enumerate}
\alphaeqn
\item[(a)] $\xi\lhd\eta$ implies $\Phi_1(\xi)\lhd\Phi_1(\eta)$ and 
\begin{equation}
\label{16a}
\Phi_1(\xi\circ\eta) = \Phi_1(\xi)\circ\Phi_1(\eta).
\end{equation}			
\item[(b)] $\xi\lhd\eta$ implies $\Phi_1(\eta)\lhd\Phi_1(\xi)$ and
\begin{equation}
\label{16b}		
\Phi_1(\xi\circ\eta) = \Phi_1(\eta)\circ\Phi_1(\xi).
\end{equation}
\reseteqn
\end{enumerate}
\item[(ii)] $\Phi_2$ is an automorphism or anti-automorphism of $\mathcal{U}$ in accordance with (i) (\emph{i.e.} $\Phi_1$ and $\Phi_2$ are either both automorphisms or both anti-automorphisms).
\item[(iii)] The following diagram is commutative. 


\begin{displaymath}
\begin{CD}
\mathcal{U}@>\Phi_2>>\mathcal{U} \\
@V{\sigma}VV   @VV{\sigma }V \\
\mathcal{T}@>>\Phi_1>\mathcal{T}
\end{CD}
\end{displaymath}
\begin{equation}\label{17}
\mbox{ \it i. e. }\Phi_1\circ\sigma = \sigma\circ\Phi_2.
\end{equation}
\end{enumerate}
Writing, for a nuclear element $\tau$ of $\mathcal{T}$, $\Phi_1(\tau)=\tau'$, $\mathcal{U}_\tau = \sigma^{-1}(\tau)$ and $\Phi_2 |\mathcal{U}_\tau = \Phi_{2\tau} $ (the restriction of the mapping $\Phi_2$ to the space $\mathcal{U}_\tau$), eq.(\ref{17}) implies that $\Phi_{2\tau}$ maps the logic $\mathcal{U}_\tau$ into the logic $\mathcal{U}_{\tau'}$.
\begin{enumerate}
\item[(iv)] For each $\tau\in\mathcal{N}(\mathcal{T})$, the mapping $\Phi_{2\tau} : \mathcal{U}_\tau\rightarrow\mathcal{U}_{\tau'}$ is an isomorphism of $\mathcal{U}_\tau$ and $\mathcal{U}_{\tau'}$ as logics~\cite{vsv}.
\end{enumerate}

Given $\Phi_1$ and $\Phi_2$ satisfying these properties, one can construct a mapping $\underline{\Phi_2} : \Omega\rightarrow\Omega$ which is an isomorphism of $\Omega$ as an orthoalgebra.
\begin{enumerate}
\item[(v)] The mappings $\Phi_3$ and $\underline{\Phi_2}$ satisfy the condition 
\begin{eqnarray}
\label{18}
\mbox{Re}\left[\Phi_3(d)\left(\underline{\Phi_2}(\underline\alpha),\underline{\Phi_2}(\underline\beta)\right)\right] = &\mbox{Re}\left[d\left(\underline\alpha,\underline\beta\right)\right] &  \\
 & \mbox{ for all } d\in\mathcal{D} & \mbox{ and all } \underline\alpha, \underline\beta \in\Omega. \nonumber
\end{eqnarray}
\end{enumerate}
The condition (\ref{18}) serves to ensure that a symmetry operation preserves the decoherence condition between histories and probabilities of histories in a decoherent set.

In~\cite{tdyj2}, symmetries in which $\Phi_1$ and $\Phi_2$ are automorphisms were called orthochronous and those in which they are anti-automorphisms non-orthochronous. A general criterion for physical equivalence of histories was formulated in terms of their transformation into each other under orthochronous symmetries. This criterion was shown to cover various notions of physical equivalence of histories considered by Gell-Mann and Hartle~\cite{mgjbh} as special cases.

The definition of symmetry given above is quite general and can be adopted in the formalism of section \ref{sectwo}. The mapping which has some special features in the present context is $\Phi_1$. (Special features will emerge in $\Phi_2$ if more structure in the space $\mathcal{U}$ is incorporated in terms of, for example, field theoretic notions. This will not be done here.) We shall, therefore, concentrate on the mapping $\Phi_1$ in the remainder of this section. 

The mapping $\Phi_1$, being an (anti-) automorphism of the partial semigroup $\mathcal{T}$, maps nuclear elements to nuclear elements in a one-to-one manner. Recall that the nuclear elements of $\mathcal{T}$ are defined as collection of basic regions [see eq.(\ref{seven})] and that our definition of basic regions allows points, spacelike curves and 2- and 3-dimensional spacelike regions to be basic regions. All the basic regions are, of course, special cases of nuclear temporal supports. Now, any mapping between two sets preserves inclusion relations between their subsets. It follows that $\Phi_1$ maps in one-to-one manner basic regions onto basic regions preserving their dimensionalities, \emph{i. e.} it maps points to points, spacelike curves to spacelike curves and two- and three-dimensional basic regions to, respectively, two- and three-dimensional basic regions. It follows, in particular, that, considering $M$ as a subset of $\mathcal{T}$, $\Phi_1$ induces an invertible mapping of $M$ onto itself (which we shall denote as $\Phi^M_1$).

In the present context, orthochronous(non-orthochronous) symmetries are those preserving(reversing) the temporal order of the basic regions; this implies that, for an orthochronous(non-orthochronous) symmetry, the mapping $\Phi_1^M$ maps past/future lightcones of spacetime points to past/future(future/past) lightcones. 

We shall prove below that $\Phi^M_1$ must be a conformal isometry of the spacetime $(M,g)$. In the proof, we shall need to invoke a continuity argument. To ensure legitimacy of such an argument, we need to define a topology on $\mathcal{T}$ so as to make it a topological partial semigroup and impose the continuity condition on $\Phi_1$. [Similar continuity requirements should be understood on $\Phi_2$ and $\Phi_3$ (with appropriate topologies defined on relevant spaces); they will, however, not be discussed here.]

The construction of a topology on $\mathcal{T}$ is described in the appendix. It is shown there that (considering $M$ as a subset of $\mathcal{T}$), the subspace topology on $M$ coincides with the manifold topology of $M$. It follows that the continuity of $\Phi_1$ implies continuity of $\Phi^M_1$ in the manifold topology.

We therefore, have an invertible continuous mapping $\Phi^M_1:M\rightarrow M$ which maps spacelike curves onto spacelike curves and \emph{ vice versa}. It follows that it maps non-spacelike curves onto non-spacelike curves and \emph{vice versa}. Now, null curves can be realized as limits of spacelike curves. Being continuous, $\Phi^M_1$ must map null curves onto null curves and \emph{vice versa} and therefore timelike curves to timelike curves and \emph{vice versa}. Considering the transformation of a small neighborhood of a point $p$ of $M$ and employing the usual local coordinate representation of line elements, we have, therefore, 
\begin{equation}
\label{19}
g'_{\mu\nu}(p')\Delta {x^\mu}^{'} \Delta {x^\nu}^{'}=\lambda_p g_{\mu\nu}(p)\Delta x^\mu \Delta x^\nu .
\end{equation}
where primes indicate transformation under $\Phi_1$ and $\lambda_p$ is a positive constant (possibly dependent on $p$). Since the point $p$ in eq.(\ref{19}) is arbitrary, the mapping $\Phi^M_1$ must be a conformal isometry of the spacetime $(M,g)$.


\section{CONCLUDING REMARKS}
\label{secfour}

1. Our treatment of quasitemporal structure for histories relating to dynamics of (closed) systems in curved spacetime is, although essentially along the lines of Isham~\cite{cji} (in the sense that it employs a partial semigroup structure defined in terms of light cones), differs from it substantially in detail. Our definition [eq.(\ref{six})] of temporal order differs from that of Isham [eq.(\ref{five})] to achieve consistency with the convention of~\cite{tdyj1}. More importantly, our basic regions are spacelike surfaces with dimensionality less than or equal to three in contrast to Isham's four-dimensional basic regions. Apart from the relative merits of the two schemes mentioned in section \ref{secone}, allowing basic regions of all dimensions less than or equal to three has the advantage that one can, for example, consider histories of systems involving both fields and particles. Moreover, as a bonus, this generality has made it possible to obtain the interesting result discussed below.

2. The main result of some interest obtained in the present work is the one obtained in the previous section, namely that a symmetry of a history theory (as formulated in~\cite{tdyj1} and~\cite{tdyj2}) relating to dynamics of systems (particles, fields, $\ldots$) in general curved spacetimes must have associated with it a transformation of the spacetime $M$ which is a conformal isometry of the underlying spacetime metric. Indeed all fundamental symmetries in various domains of physics - nonrelativistic/relativistic, classical/quantum, particles/fields/strings dynamics - satisfy this condition. It is, indeed, very satisfying that such a result should appear at the present level of generality.

We would like to recall here another instance of a general theorem~\cite{hvw} - dating more than three decades back- relating to symmetries appearing in the treatment of (quantum) dynamics of systems in terms of histories-like objects. This is what we have called `generalized Wigner theorem' in~\cite{tdyj2}. This theorem characterizes the general symmetries in quantum dynamics more comprehensively than the traditional Wigner's theorem does. Ref.~\cite{hvw} is probably the first work employing histories-like objects in the treatment of classical/quantum dynamics of systems. Unfortunately, it is rarely mentioned in the literature on history theories. We, in our papers (~\cite{tdyj1},~\cite{tdyj2}), have tried to restore some justice in this connection.

This work was supported, in part, by NSF grant DMR-9714055.


\section*{Appendix}
\subsection*{Construction of Topology for the space $\mathcal{T}$ of Temporal Supports Constructed in section 
\ref{sectwo}}

First we recall some definitions and results about topological spaces~\cite{jlk}. We shall assume that the reader is familiar with the definition of topology in terms of open sets, neighborhoods and the concept of continuity of mappings between topological spaces.

Let $(X,\mathcal{O})$ be a topological space (which means that $X$ is a nonempty set and $\mathcal{O}$ is the family of open subsets of $X$). A family $\mathcal{B}$ of subsets of $X$ is said to be a \emph{base} for the topology $\mathcal{O}$ if, for each point $x$ of $X$ and each neighborhood $V$ of $x$, there is a member $W$ of $\mathcal{B}$ such that $x\in W\subset V$. If $\mathcal{B}$ is a subfamily of $\mathcal{O}$ , it is a base for $\mathcal{O}$ if and only if each member of $\mathcal{O}$ is a union of members of $\mathcal{B}$.

A family $\mathcal{S}$ of subsets of $X$ is said to be a \emph{subbase} for the topology $\mathcal{O}$ if the family of finite intersections of members of $\mathcal{S}$ is a base for $\mathcal{O}$ (equivalently, if each member of $\mathcal{O}$ is the union of finite intersections of members of $\mathcal{S}$). According to theorem (12) of chapter (1) of~\cite{jlk}, every nonempty family $\mathcal{S}$ of subsets of a nonempty set $X$ is a subbase for some topology on $X$. This topology is uniquely defined by $\mathcal{S}$ and every member of $\mathcal{S}$ is an open set in the topology determined by $\mathcal{S}$.

According to theorem 1(c) of chapter (3) of~\cite{jlk}, a mapping $f$ of a topological space $X$ into a topological space $Y$ is continuous if and only if the inverse image under $f$ of every member of subbase for the topology on $Y$ is an open set in $X$. 

We shall construct a topology for $\mathcal{T}$ by choosing a family of subsets of $\mathcal{T}$ as a subbase. Given an element $\xi$ of $\mathcal{T}$ as in eq.(\ref{nine}) with nuclear temporal supports $\tau_i$ of the form [see eq.(\ref{seven})]
\begin{equation}
\tau_i=\{B_1^i,B_2^i,\ldots\}
\end{equation}
we introduce the collections (families of subsets of $M$)
\begin{equation}
\label{a1}
N(\tau_i)=\{N_1(B_1^i),N_2(B_2^i),\ldots\}
\end{equation}
where $N_j(B_k^i)$ is an open neighborhood of the basic region $B_k^i$ in the manifold topology of $M$; we also introduce the collection
\begin{equation}
\label{a2}
\tilde{N}(\xi) =\{\ldots,N_1(\tau_1),N_2(\tau_2),\ldots\}
\end{equation}
where each entry on the right is a collection of the form of eq.(\ref{a1}). Finally, we consider the family $\mathcal{F}$ of collections of the form (\ref{a2}) for all the elements of $\mathcal{T}$: 
\begin{equation}
\label{a3}
\mathcal{F}=\{\tilde{N}(\xi); \xi\in\mathcal{T}\}.
\end{equation}
We give a topology to the space $\mathcal{T}$ by stipulating that $\mathcal{F}$ be a subbase of that topology.

To show that, with this topology, $\mathcal{T}$ is a partial semigroup, we must show that the composition rule in $\mathcal{T}$ represented by an equation of the form $\xi\circ\eta=\zeta$ is a continuous mapping of (a subset of) the Cartesian product $\mathcal{T}\times\mathcal{T}$ into $\mathcal{T}$. To show this, it is adequate to show that the inverse image of any member of a subbase in the image space $\mathcal{T}$ is an open set in the product topology of $\mathcal{T}\times\mathcal{T}$. This is easily verified by making use of the definition of the composition rule in eq.(\ref{ten}), the construction of the topology for $\mathcal{T}$ above and the definition of the product topology.

Since points of $M$ are basic regions and, therefore, (nuclear) elements of $\mathcal{T}$, $M$ can be considered as a subset of $\mathcal{T}$. We shall now show that the subspace topology of $M$ induced by the topology of $\mathcal{T}$ constructed above coincides with the manifold topology of $M$.

Open sets of $M$ in the subspace topology are the intersections of open sets of $\mathcal{T}$ with $M$. Consider first the family $\mathcal{F}_M$ consisting of intersections of the members of the family $\mathcal{F}$ with $M$. The members $\mathcal{F}_M$ are subsets of $M$ which are open neighborhoods of points of $M$ in the manifold topology. These are also open subsets of $M$ in the subspace topology. Now, recalling the set theoretic relations
\begin{eqnarray}
\label{a4}
\left(A\cap B\right)\cap M & = &\left(A\cap M\right)\cap\left(B\cap M\right)\nonumber \\
\left(\cup_i B_i\right)\cap M & = & \cup_i\left(B_i\cap M\right) \nonumber
\end{eqnarray}
we see that the family $\mathcal{F}_M$ constitutes a subbase for the subspace topology of $M$. It is clearly also a subbase for the manifold topology of $M$. It follows that the subspace topology of $M$ coincides with its manifold topology. The choices in the initial steps in the construction of the topology for $\mathcal{T}$ above [see eq.(\ref{a1}) above] were made precisely to insure this.



\end{document}